\documentclass[aps,twocolumn,prd,nofootinbib,superscriptaddress]{revtex4}
\usepackage{amsmath}
\usepackage{graphicx}
\usepackage{dcolumn}
\usepackage{bm}
\usepackage{amssymb}
\usepackage{latexsym}

\newcommand{\be}{\begin{equation}}
\newcommand{\ee}{\end{equation}}
\newcommand{\bq}{\begin{eqnarray}}
\newcommand{\eq}{\end{eqnarray}}

\begin{document}

\title{Holographic kinetic k-essence model}

\author{Norman Cruz}
\altaffiliation{ncruz@lauca.usach.cl} \affiliation{Departamento de
F\'\i sica, Facultad de Ciencia, Universidad de Santiago de Chile,
Casilla 307, Santiago, Chile.}
\author{Pedro F. Gonz\'{a}lez-D\'\i az}
\altaffiliation{a.rozas@cfmac.csic.es} \affiliation{Colina de los
Chopos, Instituto de F\'\i sica Fundamental, Consejo Superior de
Investigaciones Cient\'\i ficas, Serrano 121, 28006, Madrid,
Spain}
\author{Alberto Rozas-Fern\'andez}
\altaffiliation{a.rozas@cfmac.csic.es} \affiliation{Colina de los
Chopos, Instituto de F\'\i sica Fundamental, Consejo Superior de
Investigaciones Cient\'\i ficas, Serrano 121, 28006, Madrid,
Spain}
\author{Guillermo S\'anchez}
\altaffiliation{gsanchez@usach.cl} \affiliation{Departamento de
Matem\'atica y Ciencia de la Computaci\'on, Facultad de Ciencia,
Universidad de Santiago de Chile, Casilla 307, Santiago, Chile.}

\date{\today}
\begin{abstract}
We consider a connection between the holographic dark energy
density and the kinetic k-essence energy density in a flat FRW
universe. With the choice $c\geq1$, the holographic dark energy
can be described by a kinetic k-essence scalar field in a certain
way. In this paper we show this kinetic k-essential description of
the holographic dark energy with $c\geq1$ and reconstruct the
kinetic k-essence function $F(X)$.
\end{abstract}

\maketitle

There is a strong belief that the universe is undergoing an epoch
of accelerated expansion. Recent cosmological observations from
Type Ia supernovae (SN Ia) [1], Cosmic Microwave Background (CMB)
anisotropies measured with the WMAP satellite [2], Large Scale
Structure [3], weak lensing [4] and the integrated Sach-Wolfe
effect [5] provide an overwhelming evidence in favour of a present
accelerating universe. Within the framework of the standard
Friedman-Robertson-Walker cosmology, this present acceleration
requires the existence of a negative pressure fluid, dubbed dark
energy, whose pressure $p_{\Lambda}$ and density $\rho_{\Lambda}$
satisfy $\omega_{\Lambda}=p_{\Lambda}/\rho_{\Lambda}<-1/3$.
Unsurprisingly, the unknown nature and origin of dark energy has
become a fundamental problem in theoretical physics and
observational cosmology. The cosmological constant (or vacuum
energy) is the most obvious candidate to address this issue as it
complies well with the cosmological tests at our disposal.
However, the well known problem of the cosmological constant and
the coincidence problem [6] are enough reasons to look for
alternatives. Interesting proposals are the quantum cosmic model
[7] and $f(R)$ theories (see [8] for recent reviews and references
therein). On the other hand, we have a plethora of dynamical dark
energy models such as quintessence [9], tachyon [10], phantom[11],
quintom [12], etc. But these scalar field dark energy models are
only seen as an effective description of the underlying theory of
dark energy.

In search of a more profound approach, holographic dark energy
models [13, 14, 15] have been recently advanced which rely on the
holographic principle [16], which is believed to be a fundamental
principle in the quantum theory of gravity. Therefore these models
incorporate significant features of the underlying theory of dark
energy. The holographic principle is a conjecture stating that all
the information stored within some volume can be described by the
physics at the boundary of the volume and, in the cosmological
context, this principle will set an upper bound on the entropy of
the universe. With the Bekenstein bound in mind, it seems to make
sense to require that for an effective quantum field theory in a
box of size $L$ with a short distance cutoff ( $UV$ cutoff:
$\Lambda$), the total entropy should satisfy the relation
\begin{equation}
L^{3}\Lambda^{3}\leq S_{BH}=\pi L^{2}M_{p}^{2},
\end{equation}
where $M_{p}$ is the reduced Planck mass and $S_{BH}$ is the
entropy of a black hole of radius $L$ which acts as a long
distance cutoff (IR cutoff: $L$). However, based on the validity
of effective quantum field theory Cohen et al [13] suggested a
more stringent bound, requiring that the total energy in a region
of size $L$ should not exceed the mass of a black hole of the same
size.

Therefore, this $UV-IR$ relationship gives an upper bound on the
zero point energy density

\begin{equation}
\rho_{\Lambda}\leq L^{-2}M_{p}^{2}
\end{equation}which means that the maximum entropy is
\begin{equation}
S_{max}\approx S^{3/4}_{BH}.
\end{equation}
The largest $L$ is chosen by saturating the bound in Eq.(2) so
that we obtain the holographic dark energy density
\begin{equation}
\rho_{\Lambda}= 3c^{2}M^{2}_{p}L^{-2}
\end{equation} where c is a free dimensionless $O(1)$ parameter and the
coefficient 3 is chosen for convenience. Interestingly, this
$\rho_{\Lambda}$ is comparable to the observed dark energy density
$\sim10^{-10}eV^{4}$ for $H=H_{0}\sim10^{-33}eV$, the Hubble
parameter at the present epoch. The fact that quantum field theory
over-counts the independent physical degrees of freedom inside the
volume explains the success of this estimation over the naive
estimate $\rho_{\Lambda}=O(M^{4}_{p})$. Therefore, holographic
dark energy models have the advantage over other models of dark
energy in that they do not need an $ad hoc$ mechanism to cancel
the $O(M^{4}_{p})$ zero point energy of the vacuum.

If we take $L$ as the Hubble scale $H^{-1}$, then the dark energy
density will be close to the observational result. However, Hsu
\cite{Hsu:2004ri} pointed out that this yields a wrong equation of
state for dark energy. This led  Li \cite{Li:2004rb} to propose
that the IR cut-off $L$ should be taken as the size of the future
event horizon of the universe
\begin{equation}
R_{\rm eh}(a)=a\int\limits_t^\infty{dt'\over
a(t')}=a\int\limits_a^\infty{da'\over Ha'^2}~.\label{eh}
\end{equation}

This allows to construct a satisfactory holographic dark energy
model that may provide natural solutions to both dark energy
problems as showed in [15]. It is worth remarking, too, that the
holographic dark energy model has been tested and constrained by
several astronomical observations [17]. If we then assume the
holographic vacuum energy scenario as the underlying theory of
dark energy, we want to see how the scalar field model can be used
to effectively describe it. Some work has been done in this
direction. Holographic quintessence and holographic quintom models
have been discussed in [18] and [19] respectively and the
holographic tachyon model in [20]. Other relevant works can be
found in [21]. Our present work aims at constructing the
holographic kinetic k-essence model of dark energy, relating the
kinetic k-essence scalar-field with the holographic dark energy.

In order to build our holographic model, we impose the holographic
nature to the kinetic k-essence, i.e., we identify $\rho_{\phi}$
with $\rho_{\Lambda}$.

We consider a universe filled with a matter component $\rho_{m}$
(including both baryons and cold dark matter) and an holographic
kinetic k-essence component $\rho_{\phi}$, the Friedman equation
reads
\begin{eqnarray}\label{VV}
3M_{P}^{2}H^{2}= \rho_{m} + \rho_{\phi},
\end{eqnarray}
or equivalently
\begin{eqnarray}  \label{Fridmann1}
H(z)= H_{0}\left ( \frac{\Omega_{m0}(1+z)^{3}}{1-\Omega_{\phi}}
\right )^{1/2}
\end{eqnarray}
where $z=(1/a)-1$ is the redshift of the universe. From the
definition of the holographic dark energy and the definition of
the future event horizon, we find
\begin{eqnarray}  \label{holo}
\int_{a}^{\infty}\frac{da^{\prime}}{Ha^{\prime
2}}=\int_{x}^{\infty}\frac{dx}{Ha}=
\frac{C}{\sqrt{\Omega_{\phi}}Ha}
\end{eqnarray}
The Friedman equation (\ref{Fridmann1}) implies
\begin{eqnarray}  \label{Ha}
\frac{1}{Ha}=\sqrt{a(1-\Omega_{\phi})} \frac{1}{H_{0}\sqrt{
\Omega_{m0}}}
\end{eqnarray}
Substituting ~(\ref{Ha}) into ~(\ref{holo}), we obtain the
following equation
\begin{eqnarray}  \label{Ha1}
\int_{x}^{\infty} e^{x'/2} \sqrt{1-\Omega_{\phi}}dx'=Ce^{x/2}
\sqrt{\frac{1}{\Omega_{\phi}}-1},
\end{eqnarray}
where $x=\ln a$. The differential equation for the fractional
density of dark energy is obtained by taking the derivative with
respect to $x$ in both sides of equation~(\ref{Ha1}), yielding
\begin{eqnarray}  \label{Omegadez}
\Omega_{\phi}^{'}=-(1+z)^{-1}\Omega_{\phi}(1-\Omega_{\phi}) \left
(1+ \frac{2}{c}\sqrt{\Omega_{\phi}}\right ),
\end{eqnarray}
where the prime denotes the derivative with respect to the
redshift $z$. This equation has an exact solution [15] and
describes the evolution of the holographic dark energy as a
function of the redshift. Since $\Omega_{\phi}^{'}$ is always
positive, the fraction of dark energy increases with time. From
the energy conservation equation of dark energy, the equation of
state of dark energy can be given [15]
\begin{eqnarray}  \label{eos}
\omega_{\phi}=-1-\frac{1}{3} \frac{d \ln \rho_{\phi}}{d \ln a}
=-\frac{1}{3} \left (1+ \frac{2}{c}\sqrt{\Omega_{\phi}}\right ).
\end{eqnarray}
Note that the formula $\rho_{\phi}={\Omega_{\phi}\over
1-\Omega_{\phi}}\rho_{\rm m}^0a^{-3}$ and the differential
equation of $\Omega_{\phi}$, Eq.(\ref{Omegadez}), are used in the
second equal sign.

Usually k-essence is defined as a scalar field $\phi$ with a
non-canonical kinetic energy associated with a lagrangian
$\mathcal{L} = -V(\phi) F(X)$. In the subsequent calculations, we
shall restrict ourselves to the simple k-essence models for which
the potential $ V=V_{0}=$ constant. We also assume that $V_{0}=1$
without any loss of generality. One reason for studying k-essence
it that it is possible to construct a particularly interesting
class of such models in which the k-essence energy density tracks
the radiation energy density during the radiation-dominated era,
but then evolves toward a constant-density dark energy component
during the matter-dominated era. Such behavior can to a certain
degree solve the coincidence problem [22].

We investigate a dark energy model described by an effective
minimally coupled scalar field with a non-canonical term. If for a
moment we neglect the part of the Lagrangian containing ordinary
matter, the general action for a k-essence field $\phi$ minimally
coupled to gravity is

\begin{equation}\label{actionKessence}
S= S_{G}+S_{\phi}=- \int d^{4}x \sqrt{-g}\left(\frac{R}{2}+
F(\phi,X)\right),
\end{equation}
where $F_k(\phi,X)$ is an arbitrary function of $\phi$ that
represents the k-essence action and $X=\frac{1}{2}\partial_{\mu}
\phi\partial^{\mu} \phi$ is the kinetic term.

We now restrict ourselves to the subclass of kinetic k-essence,
with an action independent of $\phi$
\begin{eqnarray}\label{actionKessence}
S= -\int d^{4}x \sqrt{-g} F(X).
\end{eqnarray}
We assume a Friedman-Robertson-Walker metric $ds^2 = dt^2 -
a^2(t)\, d\vec{x}^2$ (where $a(t)$ is the scale factor) and work
in units $c = \hbar = 1$. Unless stated otherwise, we consider
$\phi$ to be smooth on scales of interest so that $X = \frac{1}{2}
\dot{\phi}^2\geq0$. The energy-momentum tensor of the k-essence is
obtained by varying the action (\ref{actionKessence}) with respect
to the metric, yielding
\begin{eqnarray}\label{energy-momentKessence}
T_{\mu\nu}=  F_{X}
\partial_{\mu}\phi\partial^{\mu}\phi - g_{\mu\nu}F,
\end{eqnarray}
where the subscript $X$ denotes differentiation with respect to
$X$. Identifying (\ref{energy-momentKessence}) with the
energy-momentum tensor of a perfect fluid we have the k-essence
energy density $\rho_{\phi}$ and pressure $p_{\phi}$

\begin{eqnarray}\label{RoKessence}
\rho_{\phi}= F-2XF_{X}
\end{eqnarray}
and
\begin{eqnarray}\label{PressureKessence}
p_{\phi}=-F.
\end{eqnarray}

Throughout this paper, we will assume that the energy density is
positive so that $F-2 X F_X> 0$. The equation of state for the
k-essence fluid can be written as $p_{\phi} = w_{\phi}\rho_{\phi}$
with with $F>0$,
\begin{eqnarray}\label{wKessence}
w_{\phi} = \frac{p_{\phi}}{\rho_{\phi}}= \frac{F}{2XF_{X}-F}.
\end{eqnarray}
For a flat FRW metric, applying the Euler-Lagrange equation for
the field to the action (\ref{actionKessence}) we find the
equation of motion for k-essence field
\begin{eqnarray}\label{phiequation}
(F_{X}+2XF_{XX}) \ddot{\phi}+ 3HF_{X}\dot{\phi}=0,
\end{eqnarray}
which can be rewritten in terms of $X$ as
\begin{eqnarray}\label{phiequation1}
(F_{X}+2XF_{XX}) \dot{X}+ 6HF_{X}X=0,
\end{eqnarray}
where the dot denotes differentiation with respect to the cosmic
time and $H=\dot{a}/a$ is the Hubble parameter. If we now change
the independent variable from time $t$ to the scale factor $a$, we
obtain
\begin{eqnarray}\label{phiequation2}
(F_{X}+2XF_{XX})\,a\,\frac{dX}{da}+ 6F_{X}X=0.
\end{eqnarray}
This equation can be integrated exactly, for arbitrary $F$,
yielding
\begin{eqnarray}\label{phiequation3}
XF_{X}^{2}=ka^{-6},
\end{eqnarray}
where $k$ is a constant of integration [23]. Given a function
$F(X)$, Eq.(\ref{phiequation3}) allows us to find solutions $X(a)$
and then the other parameters of the k-essence fluid like
$\rho_{\phi}$, $p_{\phi}$ and $\omega_{\phi}$ as a function of the
scale factor, $a$.

From Eqs.(\ref{RoKessence}),(\ref{wKessence}) and (\ref{VV}), we e
can obtain the expression for $F$ as a function of the redshift z
\begin{eqnarray}\label{Fdez}
F(z)= -\rho_{\phi}\,\omega_{\phi}=-
3M_{p}^{2}\,H^{2}(z)\,\Omega_{\phi}(z)\,\omega_{\phi}(z).
\end{eqnarray}

Note that, since $\omega_{\phi}(z)<0$ , the above expression
indicates that $F$ is positive in this approach. If we demand that
the energy density be positive, Eq.(\ref{RoKessence}) implies that
$F_{X}<F/2X$. Therefore, for kinetic k-essence, $F>0$ and $F_X <
0$ imply that $w > -1$ (cf.[24] noticing the difference in the
sign convention for the energy density and the pressure). Now we
focus on the reconstruction of $F(X)$ in the redshift range
between $z=0$ and $z=1.8$ which is the current range for the
supernova data. We shall do so in the light of the holographic
dark energy with $c\geq 1$ as the future event horizon is only
well defined when $w \geq -1$ (see [15]). As an example, we plot
in Fig.1 some evolutions of the equation of state of the
holographic dark energy. We show in the plot the cases $c = 1,
1.1, 1.2$ and $1.3$. It is clear that for these cases $c\geq1$,
they always evolve in the region of $w \geq-1$.

\begin{figure}[htbp]
\begin{center}
\includegraphics[scale=0.93]{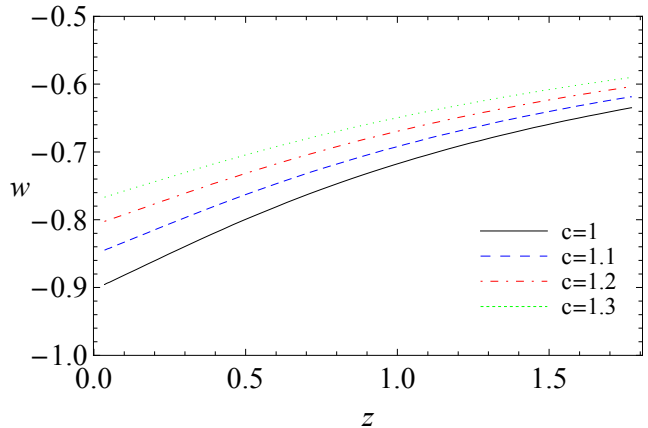}
\caption[]{\small The evolutions of the equation of state of
holographic dark energy. Here we take $\Omega_{\rm m0}=0.27$, and
show the cases for $c=1$, 1.1, 1.2 and 1.3.}\label{fig:wvsz}
\end{center}
\end{figure}

In order to carry out the numerical evaluation which allows to
find $F$ as a function of $X$, we use the dimensionless variable
$\mathcal{F}=F/(M_{p}^{2}H_{0}^{2})$. Rewriting
Eq.(\ref{phiequation3}) yields
\begin{eqnarray}\label{FF}
X \left( \frac{d\mathcal{F}}{dz}\frac{dz}{dX}\right )^{2}=
\frac{k}{(M_{p}^{2}H_{0}^{2})^{2}}\,(1+z)^{6},
\end{eqnarray}
where the usual relation $1+z=1/a$ has been used. Defining the
function $g(z)$ by
\begin{eqnarray}\label{gdez}
g(z)\equiv \left( \frac{d\mathcal{F}}{dz}\right )^{2},
\end{eqnarray}
we can obtain the following expression which allows us to
determine $X$ as a function of $z$
\begin{eqnarray}\label{Xdez}
{\int\limits_{X_{0}}^X\left(\frac{1}{M_{p}^{2}H_{0}^{2}}\right)\sqrt{\frac{k}{X'}}dX'}=
{\int\limits_0^z \frac{\sqrt{g(z)}}{(1+z)^{3}}dz}.
\end{eqnarray}
We assume that $k/X>0$ in order to have real solutions for $X$.
Integrating the above equation yields
\begin{eqnarray}\label{XdezexplicitX0}
\frac{X}{X_{0}}(z)= \left( \frac{1}{2}\,
\left(\frac{M_{p}^{2}H_{0}^{2}} {\sqrt{kX_{0}}}\right
)\,{\int\limits_0^z \frac{\sqrt{g(z)}}{(1+z)^{3}}dz} + 1 \right
)^{2},
\end{eqnarray}
which admits the following analytical solution
\begin{equation}\label{XdezexplicitX0anal}
\frac{X}{X_{0}}=\left(\frac{\Omega_{m0}\Omega_{\phi}\left(c-\sqrt{\Omega_{\phi}}\right)}{c\left(1-\Omega_{\phi}\right)
\left(\frac{3}{2}\left(\Omega_{m0}-1\right)+\Omega_{\phi0}\left(\frac{1}{2}+\frac{\sqrt{\Omega_{\phi0}}}{c}\right)\right)}\right)^{2}
\end{equation}
where $X_{0}$ and $\Omega_{m0}$ are the current values for $X$ and
$\Omega_{m}$.

From Eqs. (\ref{Fdez}) and (\ref{XdezexplicitX0anal}) we can
obtain the function $\mathcal{F}=\mathcal{F}(X/X_{0})$.

As me mention before, from Eq. (\ref{Fdez}), $F$ must be
necessarily positive and a monotonically increasing function with
$z$ within the relevant redshift range, for an accelerating
universe with holographic dark energy. This behaviour is shown in
Fig.2.
\begin{figure}[htbp]
\begin{center}
\includegraphics[scale=0.80]{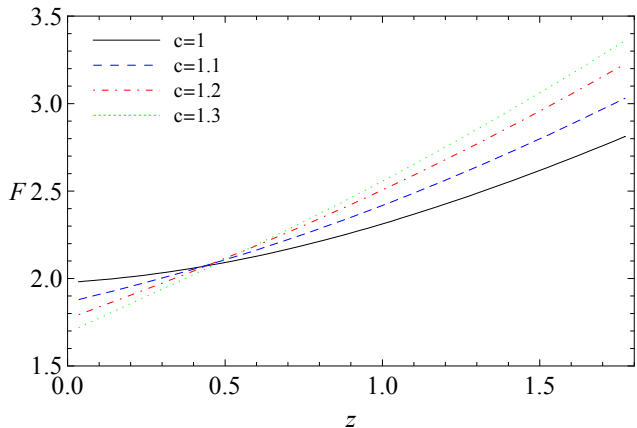}
\caption[]{\small Variation of $F(z)$ in units of
$M_{P}^{2}H_{0}^{2}$. Here we take $\Omega_{\rm m0}=0.27$, and
show the cases for $c=1$, 1.1, 1.2 and 1.3.}\label{fig:Fvsz}
\end{center}
\end{figure}

Likewise, the behaviour of $X/X_{0}$ as a function of the redshift
$z$ is showed in Fig.3.

\begin{figure}[htbp]
\begin{center}
\includegraphics[scale=0.84]{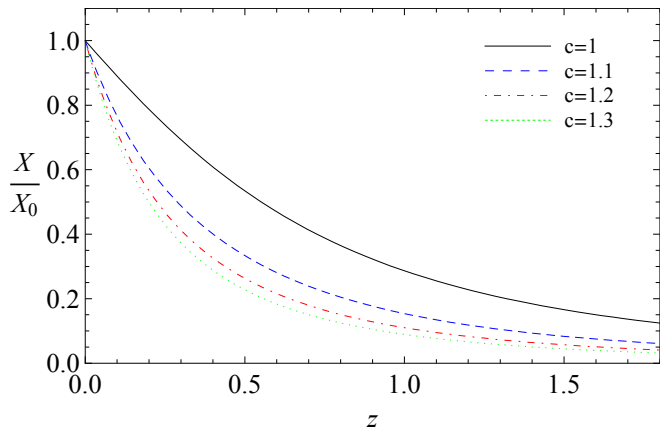}
\caption[]{\small Variation of $\frac{X}{X_{0}}(z)$. Here we take
$\Omega_{\rm m0}=0.27$, and show the cases for $c=1$, 1.1, 1.2 and
1.3.}\label{fig:Xvsz}
\end{center}
\end{figure}

The holographic kinetic k-essence, represented by the function
$\mathcal{F}$ is plotted in Fig.4 as a function of $X/X_{0}$. From
Figs. $3$ and $4$ we can see the dynamic of the k-essence field
explicitly. $F$ is a monotonically  decreasing function of $X$ in
the relevant redshift range. This is because for $X>0$, the sign
of $\frac{F_{X}}{F}$ is related to the value of $w_{\phi}$. We
should emphasise that the reconstruction of $F(X)$ only involves
the portion of it over which the field evolves to give the
requires $H(z)$. Incidentally, Figs. $2, 3$ and $4$ are very
similar to the ones shown in [25] for the transient case although
the author was dealing there with a non-holographic model in which
the ansatz for the Hubble parameter $H(z)$ was obtained by
modelling the dark energy as a Generalised Chaplygin gas. We see
that the reconstructed $\mathcal{F}=\mathcal{F}(X/X_{0})$ is a
well-behaved, single valued function.

\begin{figure}[htbp]
\begin{center}
\includegraphics[scale=0.93]{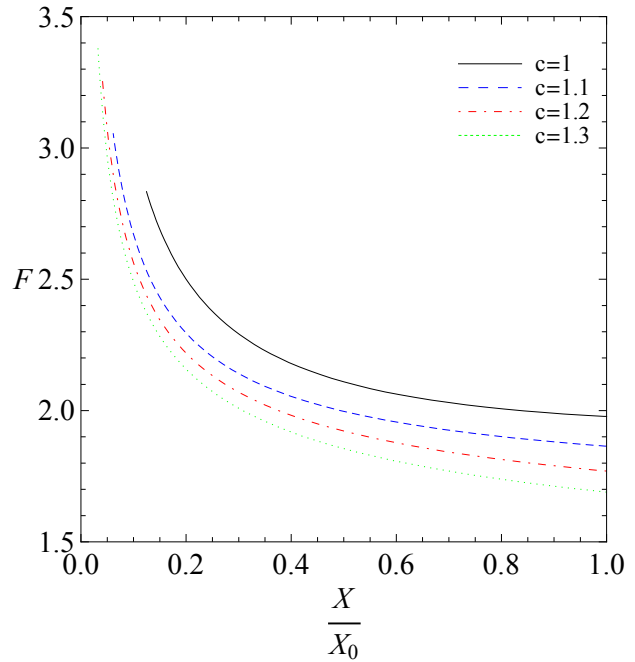}
\caption[]{\small Reconstructed $F(X/X_{0})$ in units of
$M_{P}^{2}H_{0}^{2}$. Here we take $\Omega_{\rm m0}=0.27$, and
show the cases for $c=1$, 1.1, 1.2 and 1.3.}\label{fig:FraravsXX0}
\end{center}
\end{figure}

The holographic dark energy models depend mainly on the parameter
$c$. From Eq.(\ref{eos}), we see that the equation of state
satisfies $-(1+2/c)/3\leqslant w\leqslant -1/3$ due to
$0\leqslant\Omega_{\phi}\leqslant 1$, showing that the parameter
$c$ plays a key role in the holographic evolution of the universe.

When $c\geqslant 1$, the case we are studying, the equation of
state will evolve in the region of $-1\leqslant w\leqslant -1/3$.
The value of $c$ should be determined by cosmological observations
in the holographic scenario. The case $c\geqslant 1$ is worth
investigating as current observational data cannot determine the
value of $c$ accurately. In recent fit studies, different groups
gave different values for $c$. An analysis of some of the latest
observational data, including the gold sample of 182 SNIa, the CMB
shift parameter given by the 3-year WMAP observations, and the BAO
measurement from the SDSS, showed that the possibilities of $c>1$
and $c<1$ both exist and their likelihoods are almost equal within
3 sigma error range [26].

K-essence models with different $F(X)$ have been discussed in the
literature. For the holographic kinetic k-essence model
constructed in this paper, the reconstructed $F(X)$ can be
determined from Eqs.(\ref{Fdez}) and (\ref{XdezexplicitX0anal}).
If we take $c=1$, the behaviour is similar to the cosmological
constant.

If $c>1$, the equation of state of dark energy will be always
larger than $-1$ and therefore the universe does not enter the de
Sitter phase and avoids the occurrence of a Big Rip. Thus, we see
explicitly that the value of $c$ is paramount for the holographic
dark energy model as it determines the feature of the holographic
dark energy as well as the ultimate fate of the universe.

As has been analysed above, the holographic dark energy scenario
reveals the dynamical nature of the vacuum energy. On the other
hand, as has already been mentioned, the scalar field dark energy
models are often viewed as effective description of the underlying
theory of dark energy. However, the latter theory cannot be
achieved before a complete theory of quantum gravity is
established. In spite of this, we can speculate about the
underlying theory of dark energy by taking some principles of
quantum gravity into account. The holographic dark energy model is
no doubt a tentative in this way.

To sum up, we have shown that a holographic dark energy with
$c\geq1$ can be totally described by kinetic k-essence in a
certain way. A correspondence between holographic dark energy and
kinetic k-essence has been established, and the holographic
kinetic k-essence function $F(X)$ has been reconstructed for the
redshift range between $z=0$ and $z=1.8$.

\section*{Acknowledgements}

NC acknowledges the warm hospitality of Prof. Pedro F.
Gonz\'{a}lez-D\'{i}az during the visit to the Instituto de F\'\i
sica Fundamental, Consejo Superior de Investigaciones Cient\'\i
ficas,  where part of this investigation was carried out. This
work was supported by CONICYT through grant N$^0$ 1040229 (NC),
DICYT N$^0$ 120618, of Direcci\'on de Investigaci\'on y
Desarrollo, Universidad de Santiago de Chile (GS), as well as by
DGICYT under Research Project No.~FIS2005-01180 (P.~F.~G.~D and
A.~R.~F.). A.~R.~F.~ acknowledges MEC for a FPU grant. A.~R.~F.~
benefited from discussions with David Brizuela and I\~{n}aki
Garay. 

\end{document}